\documentclass[12pt]{article}
\def\beq{\begin{equation}}
\def\eeq{\end{equation}}
\def\ap#1#2#3 {Ann. Phys. (NY) {\bf#1} (19#2) #3}
\def\err#1#2#3 {{\it Erratum} {\bf#1} (19#2) #3}
\def\ib#1#2#3 {{\it ibid.} {\bf#1} (19#2) #3}
\def\ijmp#1#2#3 {Int. J. Mod. Phys. {\bf#1} (19#2) #3}
\def\jetp#1#2#3 {JETP Lett. {\bf#1} (19#2) #3}
\def\mpl#1#2#3 {Mod. Phys. Lett. {\bf#1} (19#2) #3}
\def\np#1#2#3 {Nucl. Phys. {\bf#1} (19#2) #3}
\def\pl#1#2#3 {Phys. Lett. {\bf#1} (19#2) #3}
\def\prep#1#2#3 {Phys. Rep. {\bf#1} (19#2) #3}
\def\prev#1#2#3 {Phys. Rev. {\bf#1} (19#2) #3}
\def\prl#1#2#3 {Phys. Rev. Lett. {\bf#1} (19#2) #3}
\def\sjnp#1#2#3 {Sov. J. Nucl. Phys. {\bf#1} (19#2) #3}
\def\spj#1#2#3 {Sov. Phys. JETP {\bf#1} (19#2) #3}
\def\spu#1#2#3 {Sov. Phys. Usp. {\bf#1} (19#2) #3}
\def\zp#1#2#3 {Zeit. Phys. {\bf#1} (19#2) #3}
\def\a{\alpha}

\begin{document}
\begin{titlepage}
\begin{center}
{\Large \bf Theoretical Physics Institute \\
University of Minnesota \\}  \end{center}
\vspace{0.2in}
\begin{flushright}
TPI-MINN-02/49-T \\
UMN-TH-2123-02 \\
December 2002 \\
\end{flushright}
\vspace{0.3in}
\begin{center}
{\Large \bf  The onset of $e^+e^- \to \tau^+\tau^-$  at threshold
revisited.
\\}
\vspace{0.2in}
{\bf M.B. Voloshin  \\ }
Theoretical Physics Institute, University of Minnesota, Minneapolis,
MN
55455 \\ and \\
Institute of Theoretical and Experimental Physics, Moscow, 117259
\\[0.2in]
\end{center}

\begin{abstract}
The precise knowledge of the onset of the cross section $\sigma(e^+e^-
\to \tau^+\tau^-)$ at the threshold is necessary for improving the
accuracy of determination of the $\tau$ mass from the threshold
measurements. The QED radiative corrections of relative order $\alpha$
and $\alpha^2$, additional to the well known Coulomb factor, are
considered in the threshold region. The correction terms of order
$\alpha^2$ are calculated, which contain coefficients enhanced by large
parameters. As a result it is argued that the known  $O(\alpha)$
corrections provide the accuracy of the description of the cross section
close to $10^{-4}$, rather than $10^{-3}$ as claimed in a recent
literature. Also analytical expressions are provided for some limiting
cases of the corrections, previously calculated numerically.
\end{abstract}

\end{titlepage}

\section{Introduction}
The production in the electron-positron annihilation of slow
$\tau^+\tau^-$ pairs near the threshold provides a valuable tool for a
precision measurement of the $\tau$ lepton mass\cite{bes}. Such
measurement is aided by that the threshold onset of the cross section
starts with a finite step due to the Coulomb attraction between the
produced $\tau$ leptons\cite{mv}. Namely, when expressed in terms of the
ratio $R=\sigma(e^+e^- \to \tau^+\tau^-)/\sigma(e^+e^- \to \mu^+\mu^-)$,
the familiar `bare' threshold behavior 
\beq
R_0=v \, {3-v^2 \over 2}
\label{r0}
\eeq
in terms of the velocity $v$ of each of the produced $\tau$ leptons in
the c.m. frame, is multiplied by the Coulomb factor\footnote{$F_c$ is
also known as the Sommerfeld-Sakharov factor.}
\beq
F_c={\pi \a/v \over 1-\exp(-\pi \a/v)}~.
\label{fc}
\eeq
The product $R_0 \, F_c$ clearly has a finite limit equal to $3\pi 
\a/2$ at $v \to 0$. The factor $F_c$ sums up all the graphs with
exchange between the $\tau$ leptons of Coulomb quanta and expands in
power series in the parameter $\a/v$, rather than in powers of the QED
coupling $\a$. For this reason at the near threshold energies, where $v$
is not parametrically larger than $\a$, the Coulomb interaction has to
be taken into account exactly. The dependence of $F_c$ on $\a/v$ also
implies that the higher order corrections can arise both from the QED
radiative effects as powers of $\a$ as well as from the relativistic
expansion through terms with extra powers of $v^2$. The latter terms,
being modified by the Coulomb effects, are generally equivalent to the
ones with extra powers of $\a^2$. The leading QED radiative corrections
of order $\a$ to the discussed process were analysed in Ref.\,\cite{sv},
where it was shown that these corrections arise from two sources: a
velocity independent factor due to the form factor of the $\tau$
electromagnetic vertex at the threshold, and the Uehling-Serber
modification of the Coulomb potential due to the electron vacuum
polarization, whose effect has a quite nontrivial dependence on $v$. It
should be emphasized that the QED radiative corrections discussed here
are those which are on top of the Coulomb factor $F_c$. In other words
all the terms of the form $(\a/v)^n$ are accounted for exactly, and the
discussed corrections are the terms of the form $\a \, (\a/v)^n$: the
$O(\a)$ corrections, and $\a^2 (\a/v)^n$: the $O(\a^2)$ corrections.

Recently an attempt has been made\cite{rfp} at evaluating the $O(\a^2)$ 
corrections, which could potentially contain large factors that would
make them numerically significant. It was concluded in \cite{rfp} that
the coefficients in front of the $\a^2$ terms are not extraordinarily
large and that the $O(\a)$ radiative corrections\cite{sv} are sufficient
for describing the excitation curve for the $\tau^+ \tau^-$ pair near
the threshold with the relative precision not worse than $10^{-3}$. The
calculation in Ref.\,\cite{rfp} was based on essentially an adaptation
of the NRQCD methods used to describe the QCD effects near a heavy quark
flavor threshold (see the references in \cite{rfp}). The known results
for QCD radiative effects within these methods are based on the
$\overline{MS}$ renormalization scheme, and were used in such form in
Ref.\,\cite{rfp}. It should be noted however that in QED, unlike in QCD,
the renormalization is constrained by the simple requirement that the
asymptotic behavior of the Coulomb potential at long distances is given
by $\a/r$ with {\bf no corrections} in terms of the physical fine
structure constant $\a$, tabulated\cite{pdg} as $\a^{-1}=137.036$. This
property, inherent in the ``on shell" scheme, is lost in the 
$\overline{MS}$ scheme, in which the Coulomb potential does receive
radiative corrections at asymptotically long distances. Clearly this
effect is entirely spurious, since it goes away once the QED constant in
the $\overline{MS}$ scheme, $\a_{\overline{MS}}$, is expressed in terms
of the physical constant $\a$. However the necessity of including and
subsequently eliminating the spurious terms somewhat complicates, if not
obscures, intermediate calculations in the $\overline{MS}$ scheme. As an
illustration of a difficulty arising in the latter scheme even at the
level of the leading corrections it can be noticed that no distinction
has been made in Ref.\cite{rfp} between the electron and the muon vacuum
polarization in the Uehling-Serber type correction to the Coulomb
potential. In reality, however, these two effects result in corrections
to the cross section of completely different magnitude: the effect of
the electron loop at the threshold is an enhanced by a factor containing
$\ln (m_\tau \a/m_e)$ correction of order $\a$, while that of the muon
loop can rather be classified as an enhanced by the factor
$m_\tau/m_\mu$ correction of order $\a^2$ (or, alternatively, as an
$O(\a)$ correction, suppressed by $m_\tau \, \a/m_\mu$).
Numerically, at the threshold the former correction is almost 40 times
larger than the latter\cite{sv}.

The purpose of the present paper is to demonstrate that the calculation
of the $O(\a^2)$ radiative corrections is significantly more transparent
and simple in the ``on shell" renormalization scheme and to present an
evaluation of the terms in these corrections, enhanced by parametrically
`large' coefficients. The latter `large' coefficients include all the
factors singular in the limit $m_e \to 0$ as $\ln^2 m_e$ and $\ln m_e$,
the already mentioned correction due the muon vacuum polarization, and
also an enhanced $O(\a^2)$ effect of relativistic corrections, which is
proportional to the factor $\ln v$ (becoming $\ln \a$ at $v \ll \a$). As
will be shown, the combined effect of the enhanced $O(\a^2)$ corrections
is only about $10^{-4}$ in terms of the relative magnitude of their
contribution to the cross section, while the non-enhanced terms are
proportional to $(\a/\pi)^2 \approx 5 \cdot 10^{-6}$, thus allowing one
to argue that factually the corrections calculated in Ref.\,\cite{sv}
already provide the theoretical accuracy of $10^{-4}$ in the cross
section, rather than $10^{-3}$ as estimated in Ref.\,\cite{rfp}.  In
practical terms, this implies that the theoretical accuracy is
sufficient for a measurement of the $\tau$ mass down to at least $O(1\,
keV)$, provided that similar experimental accuracy can be achieved in
measurements at the $\tau$ threshold.

In Ref.\,\cite{sv} the effect of the $O(\a)$ correction due to the
vacuum polarization loop, both the electron and the muon, was presented
in a form of a two-dimensional integral, which then was calculated
numerically. This effect is also revisited in the present paper, and
explicit analytical expressions will be given for the electron loop
contribution in the limits $v \gg \a$ (but still $v \ll 1$), and $v \to
0$ ($v \ll \a$), as well as for the muon loop contribution, applicable
for all values of the velocity below approximately $m_\mu/m_\tau$.

\section{Types of radiative corrections to $\sigma(e^+e^- \to
\tau^+\tau^-)$.}

Generally the QED radiative corrections in the actual cross section of
the process $e^+e^- \to \tau^+\tau^-$ arise from the following
sources\cite{sv,rfp}:\\
{\it i} -- radiation from the initial electron and positron, \\
{\it ii} -- vacuum polarization in the time-like photon, \\
{\it iii} -- corrections to the spectral density $\rho
(q^2) = -{1 \over 3} \sum_X \langle 0 | j_\mu (-q) | X \rangle \langle X
|
j_\mu (q) | 0 \rangle$ of the electromagnetic current $j_\mu = ({\bar
\tau}
\, \gamma_\mu \tau)$ of the tau leptons.\\
{\it iv} -- interference between the effects {\it i} -- {\it iii} which
starts from the (relative) order $\a^2$.

The actual cross section at the `nominal' energy $W=\sqrt{s}$ in c.m. of
the electron-positron collision can thus be written in the form:
\beq
\sigma(W)=\int^W \, r(W,w)\,|1-\Pi(w)|^{-2}\, {\bar \sigma}(w) \,dw +
(interference~terms)~.
\label{se}
\eeq
The weight function $r(W,w)$ describes the radiation from the initial
state$^{\cite{kuraev}}$ and $|1-\Pi(w)|^{-2}$ is the factor for the
vacuum polarization$^{\cite{berends}}$ in the time-like photon. These
two effects are standard and are automatically accounted for in the data
analyses, while the dynamics of the final state is encoded in the cross
section ${\bar \sigma}(w) = 8 \pi^2 \alpha^2 \rho (w^2) /w^4$. The last
term in eq.(\ref{se}) arises from the graphs, where the lines of the
initial electron and positron and of the $\tau$ leptons are connected by
more than one photon propagators.

The subject of primary interest in the previous studies as well as in
the present one of the discussed process is the spectral density
$\rho(q^2)$. However before proceeding to a detailed discussion of the
QED corrections in $\rho$ few remarks are in order concerning the
leading contribution of the interference. Clearly, this contribution can
arise at the order $\a^2$ (relative to the `bare Coulomb' cross
section), and is thus within the scope of the discussion in the present
paper.

One potentially possible contribution in that order could arise from the
square of  box-type graphs, where the $\tau$ pair is produced through
two photons. However two photons produce the $\tau$ pair in a $C$-even
state. For non relativistic heavy leptons the states can be classified
in the standard (total spin) - (angular momentum) terms: $^{2S+1}L$ with
the production amplitude behaving near the threshold as $v^L$ (modulo
the Coulomb effects that `convert' powers of $v$ into powers of $\a$ at
$v \sim \a$).  The production of the $C$-even $S$-wave state $^1S_0$ by
the $e^+e^-$ is suppressed for chirality reasons by the factor $m_e$ in
the amplitude, which for all practical purposes makes it totally
negligible. The amplitude of production of the allowed by chirality
$C$-even $P$-wave states $^3P_1$ and $^3P_2$ contains an extra power of
the velocity, and thus the contribution of the box graphs to the cross
section near the threshold is additionally suppressed by the factor
$v^2$\cite{prf}. It can be noticed however that the amplitude for
production of the $^3P_2$ state actually contains a somewhat enhancing
factor $\ln v$ \cite{kks}. Thus the relative magnitude of the correction
due to the box graphs can in fact be estimated as $\delta \sigma/\sigma
\sim \a^2 \, v^2 \, \ln^2 v \sim \a^4 \ln^2 \a$, which is still far too
small.

The only other contribution of the relative order $\a^2$ from the
interference graphs can arise from the interference of the three photon
production amplitude with the `bare' amplitude mediated by one photon.
The graphical representation of this contribution is shown as a unitary
cut in the graph of Fig.1a. We are interested here in the terms of the
lowest order in the velocity, thus in the production amplitudes the
velocity can be set to zero.  One can verify that no singularity arises
from the electron propagators in the limit $m_e \to 0$ and thus that
this term does not contain enhancing factors singular at $m_e \to 0$. In
other words, it does not contain factors with powers of $\ln
(m_\tau/m_e)$. This nonsingular behavior can be seen e.g. based on the
Kinoshita-Lee-Nauenberg (KLN) theorem\cite{kinoshita,ln}. Indeed, assume
temporarily that the electric charge of the electron, $Q_e$, and  of the
$\tau$ lepton, $Q_\tau$, are independent parameters. According to the
KLN theorem there should be no infrared singular terms in the sum of the
probabilities including the emission and absorbtion of soft photons, and
this behavior is valid at arbitrary values of $Q_e$ and $Q_\tau$.
Expanding this total probability in powers of $Q_e$ and $Q_\tau$ one
finds that in the order $Q_e^4 \, Q_\tau^4$ the only contribution to
this probability in the limit $v \to 0$ comes from the unitary cuts of
the types shown in Fig.1, since at $v \to 0$ there are no cuts that
would go across the photon line as well as across the $\tau$ pair, and
also the absorbtion of soft photons by the $\tau$ leptons is irrelevant,
since at $v \to 0$ the heavy leptons neither radiate nor absorb photons.
On the other hand the cut across the three photons as shown in Fig.1b
does not contain infrared singularity in the limit $m_e \to 0$: there is
no singularity in the integrals over the energies of the photons, since
the amplitude $\tau^+\tau^- \to 3\gamma$ has no such singularity in the
photon energies\footnote{This of course is known since the work
\cite{op}.}, while the collinear singularities also do not appear in the
limit $m_e \to 0$, since there is only one electron propagator factor
per each photon.

\begin{figure}[h]
  \begin{center}
    \leavevmode
\thicklines
\unitlength 1mm
\begin{picture}(136.00,40.00)
\put(10.20,40.00){\line(2,-3){10.00}}
\put(20.20,25.00){\line(-2,-3){10.00}}
\put(28.00,25.00){\line(1,1){10.00}}
\put(38.00,35.00){\line(1,-1){10.00}}
\put(48.00,25.00){\line(-1,-1){10.00}}
\put(38.00,15.00){\line(-1,1){10.00}}
\put(66.00,40.00){\line(-2,-3){10.00}}
\put(56.00,25.00){\line(2,-3){10.00}}
\bezier{24}(20.00,25.00)(22.00,27.00)(24.00,25.00)
\bezier{24}(24.00,25.00)(26.00,23.00)(28.00,25.00)
\bezier{24}(14.00,35.00)(16.00,37.00)(18.00,35.00)
\bezier{24}(18.00,35.00)(20.00,33.00)(22.00,35.00)
\bezier{24}(22.00,35.00)(24.00,37.00)(26.00,35.00)
\bezier{24}(26.00,35.00)(28.00,33.00)(30.00,35.00)
\bezier{24}(30.00,35.00)(32.00,37.00)(34.00,35.00)
\bezier{24}(34.00,35.00)(36.00,33.00)(38.00,35.00)
\bezier{24}(48.00,25.00)(50.00,27.00)(52.00,25.00)
\bezier{24}(52.00,25.00)(54.00,23.00)(56.00,25.00)
\bezier{24}(14.00,15.00)(16.00,17.00)(18.00,15.00)
\bezier{24}(18.00,15.00)(20.00,13.00)(22.00,15.00)
\bezier{24}(22.00,15.00)(24.00,17.00)(26.00,15.00)
\bezier{24}(26.00,15.00)(28.00,13.00)(30.00,15.00)
\bezier{24}(30.00,15.00)(32.00,17.00)(34.00,15.00)
\bezier{24}(34.00,15.00)(36.00,13.00)(38.00,15.00)
\put(15.00,30.00){\makebox(0,0)[rt]{$e$}}
\put(31.00,30.00){\makebox(0,0)[rt]{$\tau$}}
\put(60.00,30.00){\makebox(0,0)[lt]{$e$}}
\put(80.20,40.00){\line(2,-3){10.00}}
\put(90.20,25.00){\line(-2,-3){10.00}}
\put(98.00,25.00){\line(1,1){10.00}}
\put(108.00,35.00){\line(1,-1){10.00}}
\put(118.00,25.00){\line(-1,-1){10.00}}
\put(108.00,15.00){\line(-1,1){10.00}}
\put(136.00,40.00){\line(-2,-3){10.00}}
\put(126.00,25.00){\line(2,-3){10.00}}
\bezier{24}(90.00,25.00)(92.00,27.00)(94.00,25.00)
\bezier{24}(94.00,25.00)(96.00,23.00)(98.00,25.00)
\bezier{24}(84.00,35.00)(86.00,37.00)(88.00,35.00)
\bezier{24}(88.00,35.00)(90.00,33.00)(92.00,35.00)
\bezier{24}(92.00,35.00)(94.00,37.00)(96.00,35.00)
\bezier{24}(96.00,35.00)(98.00,33.00)(100.00,35.00)
\bezier{24}(100.00,35.00)(102.00,37.00)(104.00,35.00)
\bezier{24}(104.00,35.00)(106.00,33.00)(108.00,35.00)
\bezier{24}(118.00,25.00)(120.00,27.00)(122.00,25.00)
\bezier{24}(122.00,25.00)(124.00,23.00)(126.00,25.00)
\bezier{24}(84.00,15.00)(86.00,17.00)(88.00,15.00)
\bezier{24}(88.00,15.00)(90.00,13.00)(92.00,15.00)
\bezier{24}(92.00,15.00)(94.00,17.00)(96.00,15.00)
\bezier{24}(96.00,15.00)(98.00,13.00)(100.00,15.00)
\bezier{24}(100.00,15.00)(102.00,17.00)(104.00,15.00)
\bezier{24}(104.00,15.00)(106.00,13.00)(108.00,15.00)
\put(85.00,30.00){\makebox(0,0)[rt]{$e$}}
\put(101.00,30.00){\makebox(0,0)[rt]{$\tau$}}
\put(130.00,30.00){\makebox(0,0)[lt]{$e$}}
\put(35.00,6.00){\makebox(0,0)[cc]{\large $a$}}
\put(105.00,6.00){\makebox(0,0)[cc]{\large $b$}}
\put(43.00,39.00){\line(0,-1){3.00}}
\put(43.00,34.00){\line(0,-1){3.00}}
\put(43.00,29.00){\line(0,-1){3.00}}
\put(43.00,24.00){\line(0,-1){3.00}}
\put(43.00,19.00){\line(0,-1){3.00}}
\put(43.00,14.00){\line(0,-1){3.00}}
\put(94.00,39.00){\line(0,-1){3.00}}
\put(94.00,34.00){\line(0,-1){3.00}}
\put(94.00,29.00){\line(0,-1){3.00}}
\put(94.00,24.00){\line(0,-1){3.00}}
\put(94.00,19.00){\line(0,-1){3.00}}
\put(94.00,14.00){\line(0,-1){3.00}}
\end{picture}
    \caption{ A representative graph ($a$) for the interference
correction in the process $e^+e^- \to \tau^+ \tau^-$. The sum of the
unitary cuts in $a$ and $b$ contains no terms singular in $m_e$ at $m_e
\to 0$ at the threshold. (Dashed lines show the unitary cuts.)}
    \label{3gam}
  \end{center}
\end{figure}
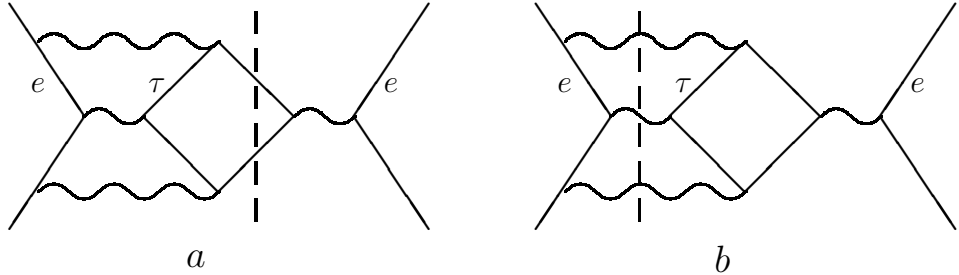

The absence of photon radiation by the $\tau$ leptons in the $v \to 0$
limit also guarantees that no terms containing $\ln v$ arise from the
graphs of the type shown in Fig.1a. Thus this contribution can only
appear as an $\a^2$ correction in the cross section with just a
numerical coefficient and containing no parametrical enhancement. As
such this contribution should be taken into account in a complete
calculation of the $\a^2$ terms, which however is beyond the intended
accuracy.

Summarizing the previous discussion, one concludes that all the QED
radiative corrections with relative magnitude of the first order in $\a$
and the parametrically enhanced ones of the second order, $\a^2$, are
contained in the corrections to the $\tau$ pair spectral density
$\rho(w^2)$, i.e. they originate from the source labeled as {\it iii}
above.

\section{First order radiative corrections. Effect of the electron
vacuum polarization}

In the lowest, `zeroth', order the effective cross section ${\bar
\sigma}(w)$ can be expressed in terms of the nonrelativistic Green's
function $G({\bf x},{\bf y},E)$ of the motion in the center of mass of
the $\tau$ lepton pair at energy $E=w-2m$\footnote{For a discussion see
Ref.\,\cite{sv} and references therein}:
\beq
{\bar \sigma}(w)= {{2 \pi^2 \alpha^2} \over {m_\tau^4}} {\rm Im}\,
G(0,0;m_\tau \, v^2)~~.
\label{nr0}
\eeq
In the absence of QED radiative effects the interaction between the
$\tau $ leptons is the Coulomb attraction, thus the Green's function is
the well known one for the Coulomb potential $V(r)=-\a/r$: $G_c({\bf
x},{\bf y},E)$. The imaginary part of the latter at ${\bf x}={\bf y}=0$
is related to that of the free-motion Green's function,
\beq
G_0({\bf x},{\bf y};{{p^2} \over m}) = {m \over {4 \pi}} {{\exp (i\,p\,
|{\bf x}- {\bf y}|)} \over {|{\bf x}- {\bf y}|}}~,
\label{g0}
\eeq 
by the Coulomb factor (\ref{fc}): ${\rm Im}\, G_c(0,0;m_\tau \, v^2) =
F_c\, {\rm Im} \,G_0(0,0;m_\tau \, v^2)$, so that the leading order
expression for  ${\bar \sigma}$ reads as
\beq
{\bar \sigma}_0={{\pi^2 \alpha^3} \over {2 m_\tau^2}}\,{1 \over {1-\exp
(-\pi \alpha/v)}}~.
\label{s0}
\eeq

At the next level of approximation, i.e. in the first order in $\a$, the
corrections to ${\bar \sigma}$ arise from two sources: from the
so-called hard correction due to a finite radiative effect in the $\tau$
electromagnetic vertex at the threshold, and from the modification of
the Coulomb interaction due to running of the coupling $\a$, which is
described by the Uehling-Serber radiative correction to the potential.   
The behavior of the Green's function at small separations ${\bf x}$ and
${\bf y}$ is determined by dynamics of the $\tau$ leptons at
characteristic distances $r_c \sim 1/p_c$, where $p_c \sim m_\tau v$ for
$v \gg \a$ and $p_c \sim m_\tau \a$ for $v \sim \a$ and $v \ll \a$. The
hard correction to the vertex comes from distances of order $1/m_\tau$,
which are thus point-like on the scale of characteristic distances in
the Green's function. Thus these two effects can be separated in terms
of eq.(\ref{nr0}) as\cite{mv79,sv}
\begin{eqnarray}
{\bar \sigma}(w)&=& {{2 \pi^2 \alpha^2} \over {m_\tau^4}}\, \left (1- {4
\, \a \over \pi} \right ) \,{\rm Im} \left[
G_c(0,0;m_\tau \, v^2)+ \delta^{(1)} G(0,0;m_\tau \, v^2) \right
]\nonumber \\
&=&{\bar \sigma_0} \, \left (1- {4 \, \a \over \pi}\right ) \,\left
(1+{2 \, \a \over 3 \pi}\, h(v) \right )~~,
\label{s1}
\end{eqnarray}
where the hard correction factor $1-4 \a/\pi$ is well known in QED (see
e.g. in the book \cite{schwinger}), and $\delta^{(1)} G$ is the
first-order correction to the Green's function due to the Uehling-Serber
correction $\delta^{(1)} V(r)$ to the Coulomb potential
\beq
\delta^{(1)} G({\bf x},{\bf y};m_\tau \, v^2)=-\int
G_c({\bf x},{\bf r};m_\tau \, v^2) \, \delta^{(1)} V(r) \, G_c({\bf
r},{\bf y};m_\tau \, v^2)\, d^3 r~,
\label{ptg}
\eeq
with $\delta^{(1)} V(r)$ given by
(see e.g. in the textbook \cite{blp}):
\beq
\delta^{(1)} V(r) =  -{{2 \alpha^2} \over {3 \pi}}{1 \over r} \,
\int_1^\infty \,
e^{-2 m_e r x} \left ( 1 + {1 \over {2x^2}}\right ){\sqrt{x^2-1} \over
{x^2}}\,dx~.
\label{us}
\eeq

The correction term $h(v)$ in eq.(\ref{s1}) due to Im$\delta^{(1)} G$
can be found by considering the modification of wave function at the
origin,
\beq
\delta^{(1)} \psi(0) = - \int G_c(0,r; m_\tau v^2)\, \delta V(r)\,
\psi_c(r) \,d^3
r ~~,
\label{pt}
\eeq
where $\psi_c(r)$ is the S-wave wave function at energy $m_\tau v^2$ in
the Coulomb field $-\alpha/r$:
\beq
\psi_c(r)=C \, e^{-i p r} \, _1 F_1(1+ i \lambda, 2 , 2i p r)
\label{psic}
\eeq
with $p=m_\tau v$, $\lambda=m_\tau \alpha /(2p)= \alpha/(2v)$ and
$C=\psi_c(0)$. Using the representation of the Coulomb Green's
function in the form
\beq
G_c(0,r; p^2/m_\tau)=-i {{m_\tau \,p} \over {2 \pi}} \, e^{i p r}\,
\int_0^\infty
e^{2 i p r t} \left ( {{1+t} \over t} \right )^{i \lambda} \, dt ~~,
\label{gfc}
\eeq
and after the integration over $r$ in eq.(\ref{pt}) the result\cite{sv}
for $h(v)$ at arbitrary $v$ is expressed in terms of a double integral:
\beq
h= - 2\lambda \, {\rm Im} \int_0^\infty
\,dt \, \int_1^\infty \, dx \left ( {{1+t} \over t} \right )^{i \lambda}
{{\left ( t+i z \, x\,v^{-1} \right )^{i \lambda -1} }
\over
{\left ( t+1+i z \, x\,v^{-1} \right )^{i \lambda +1} }}
\left ( 1 + {1 \over {2x^2}}\right ){\sqrt{x^2-1} \over
{x^2}}
\label{hf}
\eeq
with $z=m_e/m_\tau$.

The integral in eq.(\ref{hf}) can be readily calculated
numerically\cite{sv} for an arbitrary relation between $v$ and $\a$.
However it is instructive to have analytical expressions at least in the
limiting cases: $v \gg \a$ and $v \ll \a$. In either case the discussed
correction arises due to the running of the coupling $\a$ at distances
$r_c \sim p_c^{-1}$, much shorter than the electron Compton wavelength.
Thus the structure of term with $h(v)$ can be readily understood by
replacing the coupling $\a$ in the Coulomb factor $F_c$ by the effective
coupling at momentum $p_c$: $\a \to \a \,\bigl ( 1 +{2 \a \over 3 \pi}
\, \ln {p_c \over m_e} \bigr )$, thus finding
\beq
h(v)={1- (1+\pi \, \a /v)\, \exp (-\pi \, \a /v) \over 1-  \exp (-\pi \,
\a /v)} \, \ln {p_c \over m_e}~.
\label{hap}
\eeq 
In terms of this interpretation the subject of an actual calculation of
$h(v)$ reduces to finding $p_c$ in terms of $m_\tau \a$ and $m_\tau v$.
The factor in front of $\ln (p_c/m_e)$ in eq.(\ref{hap}) is obviously 
given by $(\a/F_c)\,(\partial F_c/\partial \a)$.

In each of the limiting cases the asymptotic behavior of $p_c$ is fixed
up to constants: $p_c \to const \cdot m_\tau v$ at $v \gg \a$, and $p_c
\to const \cdot m_\tau \a$ (with a different constant) at $v \to 0$. In
the former limit of large $v$ in order to find the constant one can use
eq.(\ref{ptg}) with the Coulomb Green's function replaced by the free
one (eq.(\ref{g0})), and also make use of the 
short distance limit of the Uehling-Serber correction:
\beq
\delta^{(1)}V(r)=-{\a \over r}\, \left [ 1+{2 \, \a \over 3 \pi}\, \left
( \ln {1 \over m_e \, r} - \gamma_E - {5 \over 6} \right ) \right ]
\label{uss}
\eeq
with $\gamma_E$ being the Euler's constant. The integral in
eq.(\ref{ptg}) can then be readily done, and the result reduces to the
replacement
\beq
\ln{p_c \over m_e} \to \ln{2 \, m_\tau \, v \over m_e} - {5 \over 6}
\label{lhighv}
\eeq 
in eq.(\ref{hap}) at $v \gg \a$ (but still $v \ll 1$). Taking also the
$v/\a \gg 1$ limit of the factor in front of the logarithm in
eq.(\ref{hap}), one finds the expression for $h(v)$ in this limit as
\beq
\left . h(v) \right |_{v/\a \gg1} = {\pi \, \a \over 2 \, v}\, \left (
\ln{2 \, m_\tau \, v \over m_e} - {5 \over 6} \right )~.
\label{hhighv}
\eeq
Compared to the result of the numerical integration in eq.(\ref{hf})
this expression provides the accuracy sufficient for description of the
cross section with a relative error less than $10^{-4}$ down to $v
\approx 0.04$.

One can also find analytically the behavior of $h(v)$ at $v \to 0$,
using eq.(\ref{pt}) and the explicit form of the Coulomb Green's
function at zero energy:
\beq
G_c(0,r;E \to +0)= -{m_\tau^2 \, \a \over 4 \, \sqrt{m_\tau \, \a \, r}}
\, \left [ Y_1(2 \sqrt{m_\tau \, \a \, r}) + i \, J_1(2 \sqrt{m_\tau \,
\a \, r}) \right ]~,
\label{gc0}
\eeq
where $J$ and $Y$ stand for the Bessel functions of respectively the
first and the second kind. After inserting in eq.(\ref{pt}) the
correction to the potential in the form (\ref{us}), the integral over
$r$ can be readily done, and the resulting expression for $h(0)$ reads
as
\beq
h(0)=\int_1^{\infty} \left [ 1- {m_\tau \, \a \over m_e \, x} \, \exp
\left ( - \,{m_\tau \, \a \over m_e \, x} \right ) \, K_1 \left (
{m_\tau \, \a \over m_e \, x} \right ) \right ] \, \left (1+{1 \over 2
x^2} \right ) \, {\sqrt{x^2-1} \over x^2} \, dx
\label{h0}
\eeq
with $K$ being the standard notation for the modified Bessel function of
the second kind.

The integral in eq.(\ref{h0}) can be evaluated using the presence of the
large parameter $m_\tau \, \a/m_e \approx 25.4$. The integration over
$x$ can be split in two intervals: $1 < x < X$ and $x > X$, with $X$
satisfying the conditions $1 \ll X \ll m_\tau \, \a/m_e$. After making
the obvious approximations in these two intervals one finds the result
as
\beq
h(0)=\ln {m_\tau \, \a \over m_e} + \gamma_E + {1 \over 6} + O \left [
\left ({m_e \over m_\tau \, \a} \right )^4 \right]~,
\label{h0res}
\eeq 
which in terms of eq.(\ref{hap}) corresponds to the replacement
\beq
\ln {p_c \over m_e} \to \ln {m_\tau \, \a \over m_e} + \gamma_E + {1
\over 6}~.
\label{pc0}
\eeq
Numerically, eq.(\ref{h0res}) gives $h(0)=3.980$ with a very high
accuracy, given that the subsequent terms start with the fourth power of
the small parameter $m_e/(m_\tau \a)$.

\section{Parametrically enhanced $O(\a^2)$ vertex correction.}
The effect of the muon loop arising from its contribution to the
Uehling-Serber correction to the potential can be evaluated by replacing
$m_e \to m_\mu$ in eq.(\ref{hf}) and considering there the limit $z \gg
1$. The resulting correction term $h_{(\mu)}(v)$ is then found to be
essentially constant in $v$ up to $v \approx
2m_\mu/m_\tau$\footnote{This behavior also agrees with the numerical
results of Ref.\cite{sv}}, and therefore in this region of $v$ this term
can in fact be approximated by $h_{(\mu)}(0)$ from eq.(\ref{h0}). Using
the latter expression and the fact that $m_\tau \a/m_\mu \ll 1$ one
finds
\beq
h_{(\mu)}(0)= {m_\tau \, \a \over m_\mu} \, \int_1^\infty
\left (1+{1 \over 2 x^2} \right ) \, {\sqrt{x^2-1} \over x^3} \, dx =
{9 \pi \over 32} \, {m_\tau \, \a \over m_\mu}~.
\label{hmu}
\eeq

It can be noticed that
the contribution of the muon vacuum polarization loop, as well as that
due to the heavier states, to the discussed correction for the $\tau$
pair spectral density at small velocity $v$ of the $\tau$ leptons
behaves quite differently from the contribution of the electron loop.
The reason for this difference is that a state contributing to the
absorptive part of the vacuum polarization at $q^2=s$ modifies the
interaction between the $\tau$ leptons at distances shorter than $\sim
1/\sqrt{s}$, thus the effect in the dynamics of the produced $\tau$
leptons can be considered as local, provided that the velocity satisfies
the condition $v \ll \sqrt{s}/m_\tau$. Even for the muon threshold,
$\sqrt{s}=2 m_\mu$, this condition is satisfied in all the region of
interest for the velocity. The local effect does not depend on $v$ in
this region and can be calculated by setting $v \to 0$. Therefore in
terms of the factorization formula (equations (\ref{s0}) and (\ref{s1}))
the effect of the muon loop and of higher states should rather be
written as contribution to a correction, $(\a/\pi)^2 \Delta$, in the
vertex factor:
\beq
{\bar \sigma}(w)= {{2 \pi^2 \alpha^2} \over {m_\tau^4}}\, \left ( 1- {4
\, \a \over \pi} + {\a^2 \over \pi^2}\, \Delta \right ) \,{\rm Im}
G(0,0;m_\tau \, v^2)~.
\label{s2}
\eeq
Being interpreted in terms of the correction term $\Delta$ in
eq.(\ref{s2}) the expression (\ref{hmu}) corresponds to
$\Delta_{(\mu)} = 3 \pi^2 m_\tau/(16 m_\mu)$, which although being
enhanced by the factor $m_\tau/m_\mu$, still gives a correction of only
about $1.7 \times 10^{-4}$ in eq.(\ref{s2}).

The discussed correction due to the muon loop is only a part of the
correction to the electromagnetic vertex of the $\tau$ lepton at the
threshold arising from the vacuum polarization. The latter correction
can readily be found in full by noticing that the calculation can be
formally reduced to a standard calculation of the one-loop vertex
correction with a massive photon using the dispersion relation for the
vacuum polarization $P(k^2)$ with subtraction at $k^2=0$ (the `on shell'
renormalization).
Indeed, the photon propagator with one insertion of the vacuum
polarization  can be written as
\beq
{P(k^2) \over k^2}=-{1 \over \pi} \int {{\rm Im}P(s)  \over s \,
(k^2-s)} \, ds ~,						
\label{prop}
\eeq
which clearly reduces the calculation to that
with formally a massive photon with mass $\mu=\sqrt{s}$.

At zero velocity, $v=0$, the one loop vertex correction with a massive
photon results in the following formula for the term $\Delta$ in
eq.(\ref{s2})
\beq
\Delta=-\int {{\rm Im} P(s) \over \a} \, f\left ( {\sqrt{s} \over
m_\tau} \right ) \, {ds \over s}
\label{dint}
\eeq
with the function $f(\sqrt{s}/m_\tau)$ given by
\beq
f(\xi)={2 \over 3 \xi} \, \left [ (6 + 2 \, \xi^2 + 2 \, \xi^4 - \xi^6)
\, {2 \over \sqrt{4-\xi^2}} \, \arctan {\sqrt{4-\xi^2} \over \xi} + 2 \,
\xi^5 \, \ln \xi-3 \xi - 2 \xi^3 \right ]~.
\label{frho}
\eeq

At $\sqrt{s} \ll m_\tau$ the function $f(\sqrt{s}/m_\tau)$ behaves as
\beq
f \left({\sqrt{s} \over m_\tau}\right)={2\pi \, m_\tau \over
\sqrt{s}}-4+O\left({\sqrt{s} \over m_\tau}\right)~.
\label{fir}
\eeq
Thus using the explicit expression for the muon vacuum polarization
$$-\,{{\rm Im} P(s) \over \a}={1 \over 3}\, \left (1+{2 \, m_\mu^2 \over
s} \right ) \, \sqrt{1- {4 \, m_\mu^2 \over s}}~,$$
one finds from the integral in eq.(\ref{dint}) the singular in the ratio
$m_\tau/m_\mu$ part of the muon loop contribution to $\Delta$ as
\beq
\Delta_{(\mu)}= {3 \pi^2 \, m_\tau \over 16 \, m_\mu} - {8 \over 3}\,
\ln{m_\tau \over m_\mu}~.
\label{dmufull}
\eeq
The first term here clearly reproduces the correction given by
eq.(\ref{hmu}), while the second term describes the running $\a$ effect
in the hard correction due to the fact that it comes from distances of
order $1/m_\tau$ and thus should actually be written as $-4
\a(m_\tau)/\pi$.

The singular infrared behavior of the vertex correction makes it
necessary to return to the discussion of the  electron loop effect in
the vacuum polarization contribution to the hard correction. For the
electron loop, setting $v \to 0$ in the calculation of the vertex
correction is in fact not legitimate, since $m_\tau v$ (or $m_\tau \a$)
is obviously not small as compared to $m_e$, and the leading infrared
term  has to be accounted for within the Coulomb dynamics of the $\tau$
lepton pair. This has been done in the previous section in terms of the
correction factor $h(v)$. On the other hand the contribution to $\Delta$
resulting from the subsequent term of the expansion in eq.(\ref{fir})
and  given by $\Delta_{(e)}=-(8/3)\ln(m_\tau/m_e)$ is perfectly correct,
and describes the electron loop effect when $\a(m_\tau)$ in the hard
correction is expressed in terms of the physical constant $\a$.

The contribution of the heavier than the $\mu^+ \mu^-$ pair states to
the discussed correction $\Delta$, in particular of the hadron vacuum
polarization, thus contains no large logarithmic factors. Indeed, there
is no `logarithmic range' for such masses below the $\tau$ mass, and the
effect of the states heavier than $\tau$ rapidly decreases with their
mass
($f(\xi) \approx 4 \, \ln\xi^2/(3\,\xi^2)$ at $\xi^2=s/m_\tau \gg 1$).
Thus keeping only the `large' terms one can write the final expression
for $\Delta$ as
\beq
\Delta =- {8 \over 3}\, \ln{m_\tau \over m_e}+ {3 \pi^2 \, m_\tau \over
16 \, m_\mu} - {8 \over 3}\, \ln{m_\tau \over m_\mu}
\label{dfinal}
\eeq
Numerically these three terms almost cancel each other, resulting in an
extremely small value of the vertex correction $(\a/\pi)^2 \, \Delta
\approx 1.0 \times 10^{-5}$.

\section{Parametrically enhanced $O(\a^2)$ corrections in the Green's
function.}
In this section we consider the radiative effects of the second order in
$\a$ in the Green's function in eq.(\ref{s2}). These effects arise from
the second order iteration of the first order correction to the
potential given by eq.(\ref{us}), and from the second order radiative
correction $\delta^{(2)}V(r)$ to the potential. It can be noted in
connection with the latter effect that in the on-shell scheme there is
no contribution to  $\delta^{(2)}V(r)$ due to the diagrams with exchange
of two photons between the $\tau$ leptons, and all the corrections are
given by the insertions in the single photon propagator, i.e. the two
electron loop insertion and the correction in the one loop insertion.

However, in order to find only those terms of order $\a^2$ whose
coefficients are enhanced by the second and the first power of $\ln
(p_c/m_e)$ there is no need for a detailed calculation of each these
effects, and the result can be found using the KLN theorem. Indeed, the
infrared singularity in ${\rm Im}\, G(0,0, m_\tau v^2)$ in the limit
$m_e \to 0$ should be absent, provided that the result is expressed in
terms of the effective coupling at the scale $p_c$: $\a(p_c)$. Thus all
the factors with $\ln (p_c/m_e)$ arise through expressing $\a(p_c)$ in
terms of the physical constant $\a$. Up to the single logarithmic terms
of the second order the latter expression is well known:
\beq
\a(p_c)=\a + {2 \a^2 \over 3 \pi} \, \ln {p_c \over m_e} + {4 \a^3 \over
9 \pi^2} \, \ln^2 {p_c \over m_e} + {\a^3 \over 2 \pi^2} \, \ln {p_c
\over m_e}~.
\label{apc}
\eeq
Once $p_c$ is specified in terms of $m_\tau v$ and $v/\a$ from the
results of the calculation in Section 3, the discussed logarithmically
enhanced terms can be found at any velocity in the region of interest.
Writing the effect of these terms in the form of a multiplicative factor
$[1+(\a/\pi)^2\, \Phi(v)]$ in the cross section, one finds from 
the result in eq.(\ref{h0res}) the expression for $\Phi$ the limit $v
\to 0$, where this correction is maximal,
\beq
\Phi(0) =  {4 \over 9} \ln^2 {m_\tau \a \over m_e}  + \left ( {8 \over
9} \, \gamma_E + {35 \over 54} \right ) \,  \ln {m_\tau \a \over m_e} ~.
\label{d2g}
\eeq
Numerically, the effect of this correction amounts to less than $0.5
\times 10^{-4}$.

The function $\Phi(v)$ decreases with the velocity, and at $v \gg \a$
its behavior, as can be readily found from eq.(\ref{lhighv}), is given
by
\beq
\left. \Phi(v) \right |_{v \gg \a} \approx \left ({\pi \, \a \over 2 \,
v} \right )\, \left ({4 \over 9} \ln^2 {2 m_\tau v \over m_e}  -   {13
\over 54}  \,  \ln {2 m_\tau v \over m_e}\right )~.
\label{d2highv}
\eeq

\section{Relativistic corrections.}
Due to the presence of the Coulomb parameter, $\a/v$, the relativistic
effects, which formally arise from terms of order $v^2$, get converted
into corrections of order $\a^2$. Such corrections originate from
relativistic corrections to the production vertex, and from  corrections
to the Green's function arising from the relativistic terms in the
Breit-Fermi Hamiltonian. A detailed study of these effects has been
performed\cite{hoang1,hoang2,cm} in QCD for the threshold production of
heavy quarks in $e^+e^-$ annihilation. Since the results of this study
in the order $\a^2$ are not sensitive to the renormalization scheme, one
can directly apply them to the discussed case of the $\tau$ pair
production near the threshold. The relativistic correction contains as
`large' parameter $\ln p_c/m_\tau$, and keeping only the enhanced terms,
it can be written as the following multiplicative factor in the cross
section
\beq
\left \{1+ {2 \over 3}\,\a^2 \, \left [ \ln {1 \over v}- {\rm Re}\, \Psi
\left ( - i\, {\a \over 2 \, v} \right ) \right ] \right \}~,
\label{relat}
\eeq
where $\Psi$ stands for the digamma function.

Numerically, this correction term reaches its largest value $[2 \, \a^2
\, \ln(2/\a)]/3 \approx 2.0\times 10^{-4}$ at $v=0$ and slowly decreases
with velocity (e.g. the numerical value of the correction at $v=0.1$ is
$1.7 \times 10^{-4}$).

It can be also noticed in relation to the expression in eq.(\ref{relat})
that being multiplied by the Coulomb factor (\ref{fc}) it produces a
double pole in the cross section at $v = i \, n \, \a/2$ corresponding
to the $n$-th bound $^3S_1$ level of the $\tau^+\tau^-$ pair. This
double pole correctly reproduces the Breit-Fermi relativistic correction
to the energy of the bound state\cite{hoang1}. On the other hand the
singularity of the digamma function at $v \to 0$ is canceled by the
logarithmic term. These two observations are in principle sufficient to
reproduce the structure of the discussed part of the relativistic
correction and the coefficient in front of it.

\section{Summary and concluding remarks}

We are now ready to collect all the discussed terms into one formula and
to write down the complete expression for $\sigma(e^+e^-) \to
\tau^+\tau^-$ near the threshold including all corrections of order $\a$
and all the parametrically enhanced terms of order $\a^2$:
\begin{eqnarray}
{\bar \sigma}&=&{{\pi^2 \alpha^3} \over {2 m_\tau^2}}\,{1 \over {1-\exp
(-\pi \alpha/v)}}\,
\left \{1-{4 \a \over \pi}+ {2 \a \over 3 \pi} \, h(v) \right. \nonumber
\\
&&\left.+{\a^2 \over \pi^2} \, \left \{ \Phi(v) - {8 \over 3}\,
\ln{m_\tau \over m_e} + {3 \pi^2 \, m_\tau \over 16 \, m_\mu} - {8 \over
3}\, \ln{m_\tau \over m_\mu}\right. \right.  \nonumber \\
&&\left. \left. + {2 \pi^2 \over 3}\, \left [ \ln {1 \over v}- {\rm
Re}\, \Psi \left ( - i\, {\a \over 2 \, v} \right ) \right ] - {8 \over
3} h(v) \right \} \right \}~,
\label{final}
\end{eqnarray}
where the function $h(v)$ is given by the equations (\ref{hf}),
(\ref{hhighv}), and (\ref{h0res}), and $\Phi(v)$ is defined in the
Section 5 (cf. the equations (\ref{d2g}) and (\ref{d2highv})). The last
term with $h(v)$ in the $O(\a^2)$ correction in eq.(\ref{final})
obviously arises from the interference of the first order vertex
correction with the electron loop effect in the first order (cf.
eq.(\ref{s1})).

The relative magnitude of the discussed QED radiative corrections
decreases with $v$. Thus their combined effect can be majorated by their
value at $v=0$. Writing the numerical values at $v=0$ of the individual
terms in the inner curly braces in eq.(\ref{final}), the $O(\a^2)$ term
can be evaluated as $(\a/\pi)^2 \, (8.41-21.75+31.12-7.52+36.93-10.61)=
36.58 \, (\a/\pi)^2 \approx 2.0 \times 10^{-4}$.

It can be also noticed that the $O(v^2)$ relativistic term, present in
eq.(\ref{r0}) was intentionally omitted throughout the previous
discussion as well as the total energy dependence of the cross section
arising from the timelike photon propagator producing the $\tau^+\tau^-$
pair. This is justified, as long as non enhanced $O(\a^2)$ corrections
are also ignored, in the range of the velocity where $v$ and $\a$ are
considered as being parametrically of the same order, and where the
discussed effects of the interaction between the $\tau$ leptons are most
interesting. For practical reasons however it is desirable to have a
description that interpolates with a high accuracy the cross section
between the `Coulomb' region of small $v$ and the relativistic region.
An interpolating formula of this type was suggested in Ref.\cite{sv} in
terms of the well known full expression for the cross section up to
order $\a$ (but without any summation of the Coulomb terms):
\beq
\sigma_0 (e^+\,e^- \to \tau^+ \, \tau^-) ={\frac{{2 \pi \, \alpha^2
} }{{3s}}}\, v\, (3-v^2) \, \left( 1+ {\a \over \pi} \, S(v) \right )~,
\label{sigmas}
\eeq
where $S(v)$ is given by
\begin{equation}
\label{schw}
\begin{array}{ll}
S(v)= & {\frac{1 }{v}} \left \{ (1+v^2) \left [ {\frac{{\pi^2} }{6}} +
\ln
\left( {\frac{{1+v} }{2}} \right ) \, \ln \left ( {\frac{{1+v} }{{1-v}}}
\right )+ 2\, {\rm Li}_2 \left ( {\frac{{1-v} }{{1+v}}} \right ) + 2\,
{\rm %
Li}_2 \left( {\frac{{1+v} }{2}} \right ) - \right. \right. \\  & \left.
2\,
{\rm Li}_2 \left( {\frac{{1-v} }{2}} \right ) - 4 \, {\rm Li}_2(v) +
{\rm Li}%
_2(v^2) \right ] + \left [ {\frac{{11} }{8}} (1+v^2) - 3v+ {\frac{1
}{2}} {%
\frac{{v^4} }{{(3-v^2)}}} \right ]\, \ln \left ( {\frac{{1+v} }{{1-v}}}
\right )+ \\  & \left. 6v \, \ln \left( {\frac{{1+v} }{2}} \right ) - 4v
\,
\ln v + {\frac{3 }{4}} v {\frac{{(5-3v^2)} }{{(3-v^2)}}} \right \}
\end{array}
\end{equation}
with ${\rm Li}_2(x) = -\int_0^x \ln (1-t) \, dt/t = \sum_{n=1}^\infty
x^n/n^2 $. The interpolation formula,which does include the summation of
all the Coulomb terms as well as the $O(\a)$ correction to them, has the
form
\beq
{\bar \sigma} (e^+e^- \to \tau^+\tau^-) = {\frac{{2 \pi \, \alpha^2
} }{{3s}}}\, v\, (3-v^2) \,
F_c \, \left (1+{\frac{\alpha }{\pi}} S(v) - {\frac{{\pi \, \alpha}
}{{2v}}}
+{2 \a \over 3 \pi}\, h(v)\right )~.
\label{interp}
\eeq
Near the threshold this formula correctly reproduces the Coulomb
enhancement as well as the first radiative corrections to it. At all $v$
the interpolation correctly reproduces the terms of the first order in
the expansion in $\a$. From these observations it was concluded\cite{sv}
that the higher corrections to eq.(\ref{interp}) are uniformly of order
$\a^2$ at any velocity with no enhancement by inverse powers of $v$ near
the threshold. From the calculations in the present paper it is seen
that the corrections of the relative magnitude $O(\a^2)$ in fact receive
a moderate logarithmic enhancement near the threshold which however does
not exceed $2 \times 10^{-4}$ numerically\footnote{Actually, the leading
effect of the muon loop (eq.(\ref{hmu})) was included in the analysis of
Ref.\cite{sv} in the function $h(v)$, thus the numerical correction to
that analysis is less than $10^{-4}$.}. I believe that this accuracy
should be quite sufficient for all foreseeable practical purposes.\\

\noindent
{\Large \bf Acknowledgements}

I thank A. Czarnecki, M. Shifman, and A. Vainshtein for enlightening
discussions. This work is supported in part by the DOE grant
DE-FG02-94ER40823.


\begin{thebibliography}{99}
\bibitem{bes}
J.Z. Bai {\it et.al.}, BES Collab., \prev{D53}{96}{20}.
\bibitem{mv}
M.B. Voloshin, {\it Topics in Tau Physics at a Tau-Charm Factory}, Univ.
Minnesota report TPI-MINN-89/33-T (1989) (unpublished).
\bibitem{sv}
B.H. Smith and M.B. Voloshin, \pl{B324}{94}{117}. Er. {\it ibid.} {\bf
B333} (1994) 564;\ [hep-ph/9312358]

\bibitem{rfp}
P. Ruiz-Femenia and A. Pich, Phys.Rev. {\bf D64} (2001) 053001;\
[hep-ph/0103259]

\bibitem{pdg}
K. Hagivara {\it et.al} (Particle Data Group), Phys.Rev. {\bf D66}
(2002) 010001.

\bibitem{kuraev}
E.A. Kuraev and V.S. Fadin, Yad. Fiz. {\bf 41} (1985) 733,\ [Sov. J.
Nucl.
Phys. {\bf 41} (1985) 466]
\bibitem{berends}
F.A. Berends and G.J. Komen, Phys. Lett. {\bf 63B}, 432 (1976).

\bibitem{prf}
J. Portoles, P.D. Ruiz-Femenia, Eur.Phys.J. {\bf C25} (2002) 553;\
[hep-ph/0206100]

\bibitem{kks}
J.H. K\"uhn, J. Kaplan, and E.G.O. Safiani, \np{B157}{79}{125}.
\bibitem{kinoshita}
T. Kinoshita, J.Mat.Phys. {\bf 3} (1962) 650.
\bibitem{ln}
T.D. Lee and M. Nauenberg, \prev{133B}{64}{1544}.
\bibitem{mv79}
M.B. Voloshin, \np{154}{79}{365}.
\bibitem{schwinger}
J. Schwinger, {\it Particles, Sources, and Fields}, Vol. II, Reading
1973.
\bibitem{blp}
V.B. Berestetskii, E.M. Lifshits and L.P. Pitaevskii, {\it Quantum
Electrodynamics}, Pergamon 1982.

\bibitem{op}
A. Ore and J.L. Powell, \prev{75}{49}{1696}.
\bibitem{hoang1}
A.H. Hoang \prev{D56}{97}{5851};\ [hep-ph/9704325].
\bibitem{hoang2}
A.H. Hoang \prev{D56}{97}{7276};\ [hep-ph/9703404].
\bibitem{cm}
A. Czarnecki and K. Melnikov, \prl{80}{98}{2531};\ [hep-ph/9712222].





\end{thebibliography}
\end{document}